\documentclass[aps,prd,superscriptaddress,twocolumn,showpacs,showkeys]{revtex4}
\usepackage{amssymb}
\usepackage{amsmath}
\usepackage{amsfonts}
\usepackage{graphicx, float}
\usepackage{graphicx, epsfig}
\usepackage{color}
\usepackage{enumerate}

\usepackage[colorlinks=true,
            linkcolor=red,
            urlcolor=blue,
            citecolor=blue]{hyperref}
\newcommand{\beq}{\begin{equation}}
\newcommand{\eeq}{\end{equation}}
\newcommand{\bea}{\begin{eqnarray}}
\newcommand{\eea}{\end{eqnarray}}

\begin{document}
\title{ Weak Deflection Angle of Extended Uncertainty Principle Black Holes}

\author{Yashmitha Kumaran }
\email{y.kumaran13@gmail.com}
\affiliation{Physics Department, Arts and Sciences Faculty, Eastern Mediterranean
University, Famagusta, North Cyprus via Mersin 10, Turkey.}

\author{Ali \"{O}vg\"{u}n}
\email{ali.ovgun@pucv.cl}
\homepage{https://aovgun.weebly.com} 
\affiliation{Instituto de F\'{\i}sica,
Pontificia Universidad Cat\'olica de Valpara\'{\i}so, Casilla 4950,
Valpara\'{\i}so, Chile.} 
\affiliation{Physics Department, Arts and Sciences Faculty, Eastern Mediterranean
University, Famagusta, North Cyprus via Mersin 10, Turkey.}

\begin{abstract}
In this paper, we have discussed the effects of quantum fluctuations spewed by a black hole on its deflection angle. The Gauss-Bonnet theorem (GBT) is exploited with quantum corrections through the Extended Uncertainty Principle (EUP) and the corresponding deflection angle is obtained. Moreover, we have attempted to broaden the scope of our work by subsuming the effects of plasma medium on the deflection angle as well. To demonstrate the degree of difference, the acquired results are compared with the prevailing findings.

\end{abstract}


\maketitle

\section{Introduction}
The greatest advancement in physics achieved by Einstein's theory of relativity validated its stature after 103 years, when the novelty image of a supermassive black hole was released on April 10, 2019 \cite{Akiyama:2019cqa}.
Black holes have been engrossing the scientific community since LIGO successfully detected the gravitational waves from two merging black holes \cite{Abbott:2016blz,TheLIGOScientific:2017qsa,LIGOScientific:2018mvr}. 
While subsequent attempts to enhance the detector sensitivity for further detections are underway, the Event Horizon Telescope revolutionized physics by pioneering the first look of M87* \cite{Akiyama:2019cqa}. This has marked a new era in theoretical cosmology, attracting the interest of numerous researchers towards black holes \cite{Atamurotov:2013sca,Konoplya:2019sns,Bambi:2019tjh,Shaikh:2019fpu,Abdujabbarov:2017pfw,Abdujabbarov:2016hnw}. \par

A black hole is one of the objects that gives rise to the unique phenomenon of gravitational lensing. According to Einstein, when light encounters a cluster of massive objects in its trajectory towards an observer, the cluster bends the light rays forming a gravitational lens. Galaxy clusters tend to deflect the passing light with their gravitational fields, causing distortions of the source in the background. Strong lensing produces arcs and rings such as the Einstein's ring, while weak lensing is a result of minor distortions with magnifications too small to detect, unless averaged over a number of galaxies. Although weak lensing distortions are considered futile for individual galaxies, they can distinguish between various mass distributions and can, therefore, be used to study the characteristics of the infant universe and the geometry of the cosmic web. \par

In order to utilize this subtle property of differential deflection exhibited by weak lensing, it is customary to first calculate the angle of deflection of light, which depends on the mass distribution of the lensing system. The Gauss-Bonnet theorem (GBT), introduced by Gibbons and Werner (GW), is first used to calculate the deflection angle so as to account for the optical geometry \cite{R8,R5}, using a domain outside the trajectory of light. The optical metric consists of light rays treated as spatial geodesics, inducing a topological effect \cite{Ovgun:2019wej} as well.  Then, the new GW method is applied to a variety of spacetime metric of black holes and wormholes \cite{Ovgun:2018xys,Jusufi:2017mav,Jusufi:2017vew,Sakalli:2017ewb,Jusufi:2017lsl,Ono:2017pie,Jusufi:2017hed,Jusufi:2017vta,Jusufi:2017uhh,Arakida:2017hrm,Crisnejo:2018uyn,Jusufi:2018jof,Ovgun:2018fnk,Ovgun:2018ran,Ovgun:2018prw,Ono:2018ybw,Ovgun:2018oxk,Ovgun:2018tua,Ono:2018jrv,Ovgun:2018fte,Javed:2019qyg,Javed:2019a,Javed:2019b}. Defining the Euler characteristic of the topology as $\chi$ and a Riemannian metric of the symmetric lens' manifold as $g$, the domain of the surface can be written as $(D, \chi, g)$. For Gaussian curvature, $K$ and geodesics curvature, $\kappa$, GBT is formulated as \cite{R8}:
\begin{equation}
    \int \int_D K dS + \int_{\partial D} \kappa dt + \sum_i \alpha_i = 2\pi \chi(D) \label{GBT1}
\end{equation}
where, $\alpha_i$ is the exterior angle with $i$\textsuperscript{th} vertex. \newline
The deflection angle was calculated by Gibbons and Werner \cite{R8} through this method for a Schwarzschild black hole. When the source and the observer are asymptotically flat, the deflection angle is presumed to be very small. Thus, the asymptotic deflection angle, is given by \cite{R8}:
\begin{equation}
    \hat{\alpha} = - \int \int_D K dS. \label{GBT2}
\end{equation}

One of the major conundrums associated with black holes is the information paradox. Quantum mechanics predicts that the information (or the specific state) of a particle plummeting into a black hole cannot be lost; more precisely, the quantum wave function that acts as the fingerprint of the falling particle is always preserved on the surface of the black hole for billions of years. In 1974, Stephen Hawking proposed that the black holes evaporate over time releasing their mass and energy back into the universe in the form of black body radiation, known as the Hawking radiation \cite{Hawking:1974rv,Hawking:1974sw}. This arises from the fact that the black holes have temperatures, and implies that the information of the engulfed object vanishes with the black hole. Clearly, quantum mechanics and general relativity contradict each other in this case, leading to the information paradox.\par

Out of the few conjectures that solve the information paradox, the most plausible explanation was given by Hawking. He suggested that the information of the falling objects could escape from being absorbed into the black hole, by leaking out of its radiation field through quantum fluctuations. The outgoing radiation returns to the universe with the particle (perhaps, distorted) information imprinted on it. These quantum perturbations in the event horizon require quantum gravity corrections so that their spacetimes are consistent. Such quantum effects infer that the black holes obey quantum mechanics. \par

In this article, our aim is to employ this approach to determine the quantum gravity effects on the deflection angle. This paper is organized as follows: in section 2, we briefly review the spacetime described by an EUP-corrected black hole. In section 3, we calculate the deflection angle by EUP black hole using the Gauss-Bonnet theorem in weak field regions. In section 4, we extend our studies for the deflection of light by EUP black hole in a plasma medium. We conclude our results in section 5. Natural units are used throughout this paper: $G=\hbar=c=1$.

\section{An Extended Uncertainty Principle Black hole spacetime}
Consider the hydrogen atom. The kinetic energy of the electron balances out its negative potential energy, thereby preventing the electron from collapsing into the nucleus. Heisenberg stated this through the Uncertainty Principle. Since they are now recognized as quantum objects, Ronald J. Adler argues that the same idea can be applied to black holes: the Generalized Uncertainty Principle \cite{Adler:2001vs} might prevent a black hole from evaporating. Mathematically, the Heisenberg relation contains an extra term proportional to the square of momentum uncertainty. Alternatively, the Heisenberg relation can be modified with an extra term proportional to the square of position uncertainty, yielding the Extended Uncertainty Principle (EUP). If the contribution from the EUP corrections are adequate, they can be used on a large scale to compute dark matter effects, properties of the black hole, size of its photosphere, etc. \par

For a fundamental distance scale, $L$, the Heisenberg relation with the EUP correction for position uncertainty \cite{Mureika:2018gxl} can be written as:
\begin{equation}
    \Delta x \Delta p \geq 1 + \alpha \frac{\Delta x^2}{L^2},
\end{equation}
where $\alpha$ is a coupling constant. \par

It is emphasized that the uncertainty principle remains retrievable due to the condition $L \gg \Delta x$. It also implies that \cite{Bolen:2004sq} the effects of quantum gravity manifest themselves over large distances, hence validating the notion of quantum effects on macroscopic scales. \par

The line element of a spherically symmetric black hole with mass, M, subjected to the EUP correction is defined \cite{euplensing} as:
\begin{equation}
    \label{le} 
    ds^2 = - f(r) dt^2 + f(r)^{-1} dr^2 + r^2 (d\theta^2 + \sin^2 \theta d\phi^2),
\end{equation}
where:
\begin{equation}
    f(r) \equiv 1- \frac{2M}{r} \left(1+\frac{4\alpha M^2}{L^2} \right).
\end{equation}
The corresponding expression \cite{Virbhadra:1998dy} for the deflection angle of a photon verging on this black hole with the distance of closest approach, $r_0$ is written as:
\begin{equation}
    \hat{\alpha}(r_0) = -\pi + 2 \int_{r_0}^\infty dr \frac{\sqrt{f(r)^{-1}}}{r \sqrt{\frac{r^2}{r_0^2}\frac{f(r_0)}{f(r)}-1}}.
\end{equation}
If $M/r_0 \ll 1$, the deflection angle is too small: this signifies weak lensing. Subsequently, $\alpha$ increases as $r_0$ approaches the photosphere until it diverges, to produce strong lensing.

\section{Weak Deflection angle and Gauss-Bonnet Theorem}
Taking $\theta=\pi/2$ for equatorial plane, equation \eqref{le} reduces to the optical metric for null geodesics:
\begin{equation}
    \label{om} 
    dt^2 = \frac{dr^2}{f(r)^2} + \frac{r^2}{f(r)} d\phi^2.
\end{equation}
Following the computation of non-zero Christoffel symbols, the Gaussian curvature (proportional to the Ricci scalar) can be written as:
\begin{equation}
    \label{gc}
    K = \frac{R}{2} \approx - {\frac {8{M}^{3}\alpha}{{r}^{3}{L}^{2}}} - {\frac {2M}{{r}^{3}}} +\mathcal{O}(M^{4}).
\end{equation}
Using the straight line approximation \cite{R8} as $r = u/ \sin \phi$, where $u$ is the impact parameter, equation \eqref{GBT2} of GBT  suggests that:
\begin{equation}
     \label{allim}
     \hat{\alpha} = - \int_0^\pi \int_{\frac{u}{\sin \phi}}^\infty K dS,
\end{equation}
where $dS=rdr d\phi$.
Ignoring the higher order terms, equations \eqref{gc} and \eqref{allim} simplifies to the following expression for the deflection angle due to weak lensing with EUP corrections:
\begin{equation}
    \hat{\alpha}_{w} = {\frac {4M}{u}}+{\frac {16{M}^{3}\alpha}{u{L}^{2}}} \label{alpha}
\end{equation}
for weak-field limits. Thus, the EUP parameter, $\alpha$ increases the deflection angle, and deflection angle reduces to case of Schwarzschild black hole when $\alpha=0$. The deflection angle in the leading order terms is seen to be in agreement with \cite{euplensing}.

\section{Weak Deflection angle in a plasma medium}
In order to incorporate the effects of plasma, consider the case when light travels from vacuum to a hot, ionized gas medium. Let $v$ be the velocity of light through the plasma. The refractive index, $n(r)$ is described by:
\begin{equation}
    n(r) \equiv \frac{c}{v} = \frac{1}{dr/dt} \quad \quad\quad\quad\quad \{\because c=1\}
\end{equation}
The refractive index $n(r)$ for an EUP-corrected black hole is obtained as \cite{Crisnejo:2018uyn},
\begin{equation}
    n(r)=\sqrt{1-\frac{\omega_e^2}{{\omega_\infty^2}} \left[1- \frac{2M}{r} \left(1+\frac{4\alpha M^2}{L^2} \right)\right]},
\end{equation}
where $\omega_{e}$ is the electron plasma frequency and $\omega_{\infty}$ is the photon frequency measured by an observer at infinity. The line element in equation (\ref{le}) can be re-written as:
\begin{widetext}
\begin{equation}
    d \sigma ^ { 2 } = g _ { i j } ^ { \mathrm { opt } } d x ^ { i } d x ^ { j } = \frac { n ^ { 2 } ( r ) } { 1- \frac{2M}{r} \left(1+\frac{4\alpha M^2}{L^2} \right)} \left[ \frac{d r ^ { 2 }}{1- \frac{2M}{r} \left(1+\frac{4\alpha M^2}{L^2} \right)} + r^2 d \phi ^ { 2 }\right].
\end{equation}
\end{widetext}
Therefore, the optical Gaussian curvature turns out to be:

\begin{eqnarray} 
    K \approx -{\frac {2M}{{r}^{3}}}+{\frac {3{M}^{2}}{{r}^{4}}}-{\frac  {8{M}^{3}\alpha}{{L}^{2}{r}^{3}}}+  -{\frac {3M\omega_e^2}{{\omega_\infty^2}r^3}}+{\frac {12{M}^{2}\omega_e^2}{{\omega_\infty^2}{r}^{4}}} \notag \\ -\left( {\frac {12\alpha}{{\omega_\infty^2}{L}^{2}{r}^{3}}}+ {\frac {12}{{\omega_\infty^2}{r}^{5}}} \right) {M}^{3}  {\omega_e^2}.
\end{eqnarray}

\begin{figure}
\centering
\includegraphics[width=10cm]{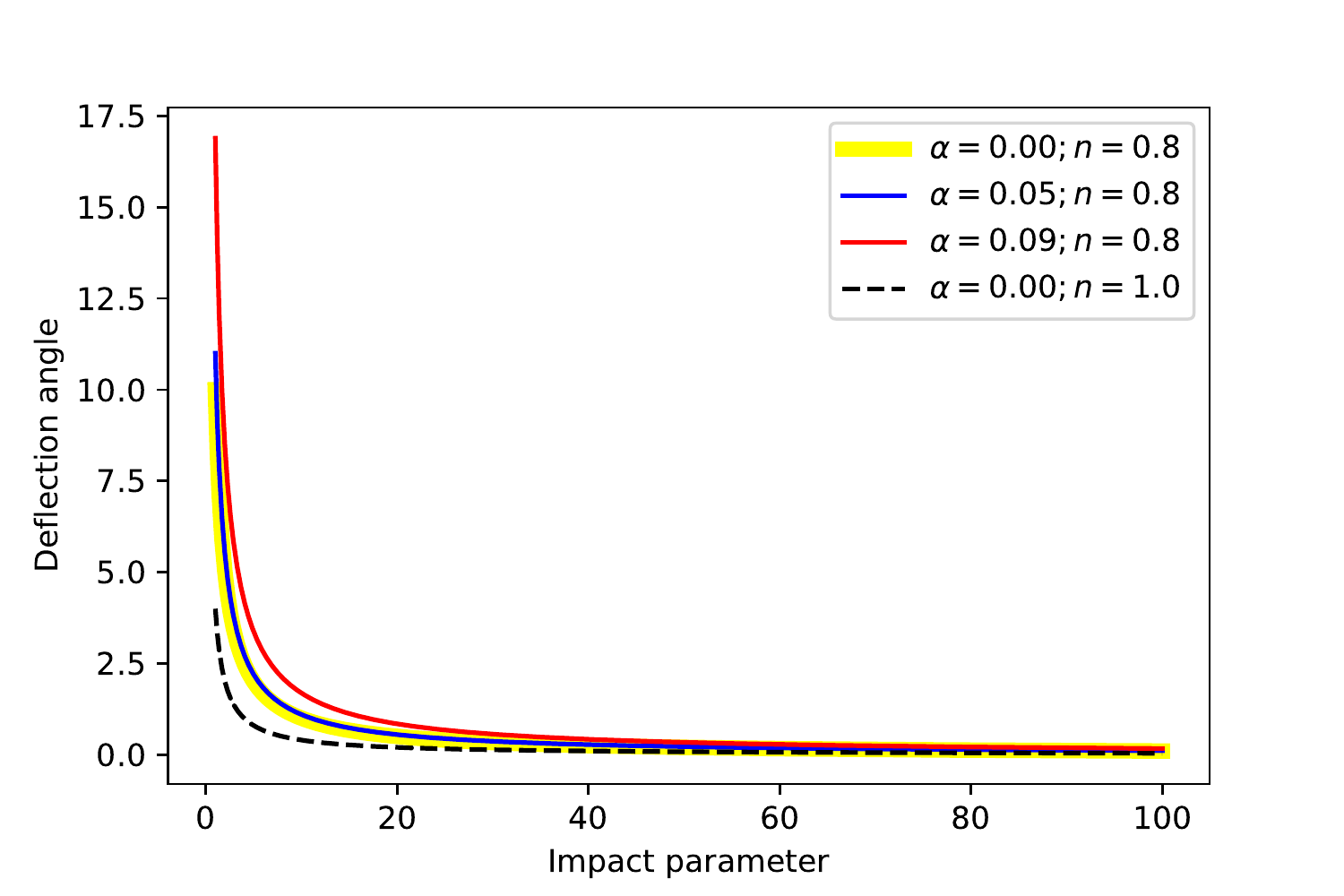}
\caption{\label{ipda} Deflection angle vs impact parameter for various values of coupling constant, $\alpha$, with refractive index, $n=0.8$.}
\end{figure}

Consequently, the deflection angle becomes:
\begin{equation}
    \hat{\alpha}_{w_{n}} \approx {\frac {4M}{u}}+{\frac {6M{\omega_e^2}}{u{\omega_\infty^2}}}+{\frac {16\alpha{M}^{3}}{u{L}^{2}}}+{\frac {24\alpha{L}^{2}{\omega_e^2}{M}^{3}}{u{\omega_\infty^2}}}\label{pal}.
\end{equation}
In an attempt to comprehend the variation indicated by equation \eqref{pal} graphically, the deflection angle was plotted against the impact parameter to obtain fig.\ref{ipda}. \newline \newline

In Fig.\ref{ipda}, the deflection angle is observed to be affected by the quantum effects for low values of the impact parameter. Evidently, the contribution of these effects (characterized by the EUP parameter, $\alpha$) is not negligible. Thus, it can be seen that the deflection angle increases when the photon rays move through a medium of homogeneous plasma. The first case - in which the EUP corrections are not considered - shows a hefty rise in the deflection angle for a plasma medium, from the typical $\hat{\alpha}_w$ that corresponds to vacuum. Although, for high impact parameter, all the behaviours seem to be in close agreement. Furthermore, the results show that as $\omega_e/\omega_\infty\to 0$, equation \eqref{pal} reduces to equation \eqref{alpha}, removing the influence of plasma. This signifies the standard case and is represented by the dashed line in the above plot.

\section{Conclusions}
In this article, we have investigated the quantum gravity effects in the vicinity of a black hole, so as to preserve the specific state of an object falling into it, hence, solving the information paradox. These quantum effects, in the presence of a plasma medium, are found to engender substantial changes in the deflection angle. This variation can be utilized to develop the precision of differential deflection established by the typical assumptions of weak lensing. The nature of equation \eqref{pal} to adapt between vacuum and plasma advocates flexibility to modify the result for a wide range of analyses. Finally, it is concluded that the EUP corrections are indeed high enough to determine various parameters on a large scale.

\acknowledgments
This work was supported by Comisi{\'o}n Nacional de Ciencias y Tecnolog{\'i}a of Chile through FONDECYT Grant $N^\mathrm{o}$ 3170035 (A. {\"O}.).

\end{document}